\documentstyle[epsfig]{l-aa}

\newcommand{\be}{\begin{equation}}
\newcommand{\ee}{\end{equation}}

\begin{document}

\thesaurus{13.07.1, 02.19.1, 09.10.1}
\title{The influence of inverse Compton scattering on GRB afterglows: one possible way to flatten and steepen 
         the light curves}
\author{D.M. Wei\inst{1,2}, 	and 	T. Lu\inst{3,4}}
\institute{Purple Mountain Observatory, Chinese Academy of Sciences, Nanjing, China
\and
	  National Astronomical Observatories, Chinese Academy of Sciences, China
\and
	   Department of Astronomy, Nanjing University, Nanjing, 210093, China
\and
        Laboratory for Cosmic-Ray and High-Energy Astrophysics, Institute of High Energy Physics,
        Chinese Academy of Sciences, Beijing, 100039, China}

\date{Received date/ Accepted date}
\maketitle
\markboth{Wei \& Lu:\,\, ICS on GRB afterglow}{}

\begin{abstract}
The fireball model of gamma-ray bursts predicted that when the energetic blast wave encountered the 
surrounding medium, there will be afterglow emission, and the subsequent afterglow observations
appeared to confirm this prediction. In this simplest fireball model, the electrons have been accelerated 
to a power law energy distribution in a relativistic blast wave, then they give afterglow emission through
synchrotron radiation. Up to now synchrotron radiation is believed to be the main mechanism of GRB
emission, however, here we will show that under some circumstances, the inverse Compton scattering
(ICS) may play an important role, and can change the light curves of GRB afterglows. Here we investigate
the effects of ICS in the relativistic case (the surrounding medium density $\rho \propto r^{-2}$) and in
the non-relativistic case (for both $\rho =constant$ and $\rho \propto r^{-2}$), we find that in the 
relativistic case the effect of ICS is usually important, while in the non-relativistic case, this effect 
is usually unimportant, unless the surrounding medium density is very high.  We show that if ICS
is important, then it can flatten and steepen the light curves of GRB afterglows, and this may provide
the explanation for some afterglow observations. 

\end{abstract}

\section{Introduction}
The observed properties of GRB afterglows are in approximate accord with the models based on
relativistic blast wave at cosmological distances. In this standard fireball model, the huge energy
released by an explosion is converted into the kinetic energy of a shell expanding at ultra-relativistic
speed. After the main GRB event has occured, the fireball continues to propagate into the surrounding
medium. The expanding shock continuously heats fresh gas and accelerates electrons to very 
high energy, which produce the afterglow emission through synchrotron radiation (Meszaros \&
Rees 1997; Vietri 1997; Waxman 1997a, 1997b; Wijers, Rees \& Meszaros 1997; Wei \& Lu 1998a).

In the simplest models, the surrounding medium density is assumed to be constant. However,
there has been increasing evidence that at least some GRBs have massive star progenitors, as 
suggested by the link between some GRBs and supernovae, which means that the GRB blast wave
should be expanding into the stellar wind of the progenitor star, the densiy $\rho \propto r^{-2}$
(Dai \& Lu 1998; Chevalier \& Li 1999a, 1999b). Chevalier \& Li (1999a, 1999b) have shown that in this 
wind density  the afterglow light curves should be steeper than that in the constant density.

Although, in principle, the standard fireball model can approximately explain the afterglow light 
curves well, there are still some problems that cannot be explained by this simplest model. For 
example, it is well known that in some GRB afterglows, the light curves cannot be described by
a simple power-law, but show sharp breaks (e.g. for GRB990123, see Castro-Tirado et al. 1999;
Kulkarni et al. 1999; Fruchter et a. 1999; Galama et al. 1999; for GRB990510, see Stanek et al.
1999; Harrison 1999; and for GRB000301c, see Rhoads \& Fruchter 2000; Masetti et al. 2000).
These observed breaks have generally been interpreted as evidence for collimation of the
GRB ejecta (Rhoads 1999), but a difficulty with this model is that the predicted break is
quite smooth , while the observed breaks are rather sharp (Panaitescu \& Meszaros 1999;
Moderski, Sikora \& Bulik 2000; Kumar \& Panaitescu 2000; Wei \& Lu 1999a). The transition 
of blast wave from relativistic to non-relativistic regime has been proposed as another 
mechanism for light curve breaks (Dai \& Lu 1999).

We note that, in the fireball model, the synchrotron radiation has been regarded as the main
mechanism for GRB emission, while the effect of inverse Compton scattering has been
neglected. However, we have shown earlier that, when the relativistic blast wave passing
through the constant surrounding medium, the ICS may have an important effect on the
emission spectrum and on the temporal behavior of afterglows (Wei \& Lu 1998b). Here
we extend our discussion to a more general case, the surrounding medium density can
be not uniform, $\rho \propto r^{-2}$, and the blast wave can be in the non-relativistic stage,
here we only consider the case that the electrons are slow cooling, which pertains to
most of the afterglow phase.
We find that, in the relativistic case, the effect of ICS is usually important, which can
flatten and steepen the afterglow light curves, while in the non-relativistic case, the effect 
of ICS is usually not important unless the surrounding medium density is very high. 
In next section, we investigate the effect of ICS in the relativistic case for wind density
$\rho \propto r^{-2}$. In section 3, we discuss the situation that the blast wave is non-relativistic
and the surrounding medium density is both uniform and non-uniform. Finally some 
discussion and conclusions are given in section 4.

\section{The effect of ICS in the relativistic case}
The process of inverse Compton scattering has been discussed by several authors (e.g.
Waxman 1997a; Dermer, Boettcher \& Chiang 1999; Panaitescu \& Kumar 2000; 
Sari \& Esin 2000), and here we will consider the influence of ICS on GRB afterglow in details.
There is a simple way to estimate
the intensity of ICS, i.e. by calculating the ratio of the synchrotron radiation energy density 
($u_{\rm syn}'$) to the magnetic energy density ($u_{\rm B}'$), $R=u_{\rm syn}'/u_{\rm B}'$, where $u_{\rm B}'=
B'^{2}/8\pi$, and $u_{\rm syn}'\sim n_{\rm e}'P_{\rm syn}'\tau' $, $n_{\rm e}'$ is the electron number density
in the comoving frame, $n_{\rm e}'=\frac{\hat \gamma \Gamma +1}{\hat \gamma -1}n$, $\Gamma $
is the bulk Lorentz factor of the blast wave, $\hat \gamma $ is the adiabatic index, and 
$\hat \gamma =4/3$ for relativistic case and $\hat \gamma =5/3$ for non-relativistic case,
$n$ is the surrounding medium density, $P_{\rm syn}'$ is the synchrotron radiation power of
a single electron, and $\tau' \sim r/\Gamma c$. The value of $R$ indicates which process
is more efficient for electron energy loss. If $R>1$, then  the effect of ICS is
important. It is easy to show that 
\be
R=10^{-24}\bar {\gamma ^{2}}r\Gamma ^{-1}\frac{\hat \gamma \Gamma +1}{\hat \gamma -1}n
\ee
where $r$ is the distance from the burst source, $\bar {\gamma ^{2}}$ is the average value
of electron Lorentz factor square, 
$\gamma _{\rm e}$ is the minimum electron Lorentz factor, $\gamma _{\rm e}=\xi _{\rm e}(\Gamma -1)
\frac{m_{\rm p}}{m_{\rm e}}\frac{p-2}{p-1}$, $\xi _{\rm e}$ is the energy fraction 
occupied by electrons, and $p$ is the index of electron distribution, for $p=3$, we have
$\gamma _{\rm e}\simeq 900\xi _{\rm e}(\Gamma -1)$, and $\bar {\gamma ^{2}}\simeq 10\gamma _{\rm e}^{2}$. This 
formula is valid for both the relativistic and non-relativistic cases, now we first consider the 
relativistic case. 

We have shown earlier that in the relativistic case and for uniform density ($n\sim 1\,{\rm cm^{-3}}$),
the effect of ICS is usually important, the value $R=48(\frac{\xi _{\rm e}^{2}}{0.1})n_{1}^{1/2}E_{52}^{1/2}
t_{\rm day}^{-1/2}$ (Wei \& Lu 1998b). Now we extend to the case for non-uniform density, $\rho \propto r^{-2}$.

Chevalier \& Li (1999a,b) have discussed the blast wave dynamical evolution in the wind environment,
and they gave $\gamma=4.2(\frac{1+z}{2})^{1/4}E_{52}^{1/4}A_{\star }^{-1/4}t_{\rm day}^{-1/4}$,
$r=2.9\times 10^{17}(\frac{1+z}{2})^{-1/2}E_{52}^{1/2}A_{\star }^{-1/2}t_{\rm day}^{1/2}\,\,{\rm cm}$, where
$A=\dot{M_{\rm w}}/4\pi V_{\rm w}=5\times 10^{11}A_{\star}\,{\rm gcm^{-1}}$, $\dot{M_{\rm w}}$ is the mass
loss rate, and $V_{\rm w}$ is the wind velocity, the reference value of $A$ corresponds to
$\dot{M_{\rm w}}=1\times 10^{-5}M_{\odot }\,\,{\rm yr^{-1}}$ and $V_{\rm w}=1000\,\,{\rm kms^{-1}}$. Thus we obtain
\be
R=54(\frac{\xi _{\rm e}^{2}}{0.1})(\frac{1+z}{2})A_{\star}t_{\rm day}^{-1}
\ee
We see that the emission power of ICS is not  smaller than that of synchrotron radiation,
thus, the effect of ICS should not be neglected.

Now let us consider the effect of ICS on GRB afterglow. The GRB emission spectrum should
consist of two components, below the critical frequency $\nu _{\rm c}$  the spectrum  is dominated
by synchrotron radiation, and above $\nu _{\rm c}$ the spectrum is dominated by ICS. We assume
that in the comoving frame the synchrotron radiation intensity has the form $I_{\nu }\propto 
\nu ^{-\alpha }$ for $\nu <\nu _{\rm m}$ and $I_{\nu }\propto \nu ^{-\beta }$ for $\nu >\nu _{\rm m}$. Since
in our situation the soft photons produced through synchrotron radiation are scattered by
the same electrons, the Compton-scattered spectrum should have nearly the same form as
that of synchrotron radiation. Therefore the total emission intensity is $I_{\nu }\propto 
\nu ^{-\alpha }$ for $\nu <\nu _{\rm m}$ or $\nu _{\rm c}<\nu <\nu _{\rm n}$, and $I_{\nu }\propto 
\nu ^{-\beta }$ for $\nu _{\rm m}<\nu <\nu _{\rm c}$ or $\nu >\nu _{\rm n}$, where $\nu _{\rm n}=
\gamma _{\rm e}^{2}\nu _{\rm m}$ 
is the peak frequency of the ICS spectrum, and $\nu _{\rm n}I_{\nu _{\rm n}}=R\nu _{\rm m}I_{\nu _{\rm m}}$. 
Then, from the relation $I_{\nu _{\rm m}}(\nu _{\rm c}/\nu _{\rm m})^{-\beta }
=I_{\nu _{\rm n}} (\nu _{\rm c}/\nu _{\rm n})^{-\alpha }$ we can obtain 
\be
\frac{\nu _{\rm c}}{\nu _{\rm m}}=a_{1}
\xi _{\rm e}^{-\frac{2\alpha }{\beta -\alpha }}(\frac{1+z}{2})^{-\frac{1+\alpha }{2(\beta -\alpha )}}
A_{\star}^{-\frac{3-\alpha }{2(\beta -\alpha )}}E_{52}^{\frac{1-\alpha }{2(\beta -\alpha )}}
t_{\rm day}^{\frac{1+\alpha }{2(\beta -\alpha )}}
\ee
where $a_{1}=540^{-\frac{1}{\beta -\alpha }}(900\times 4.2)^{\frac{2(1-\alpha )}{\beta -\alpha }}$.
So this critical frequency is dependent on the fireball parameters, i.e. the fireball energy,
surrounding gas density, energy fractions in electrons and magnetic field, and the spectrum
index of synchrotron radiation. As an example, we take $\alpha =0,\,\,\beta =1$ (these valus
are typical for observed GRB spectra), then the value of $\nu _{\rm c}$ is given by
\be
\nu _{\rm c}=10^{4}(\frac{\xi _{\rm e}^{2}}{0.1})(\frac{\xi _{\rm B}}{0.1})^{1/2}A_{\star}^{-3/2}
E_{52}t_{rm day}^{-1} \,\,\,\,{\rm eV}
\ee
where  we have taken the peak frequency of synchrotron radiation $\nu _{\rm m}=
0.4(\frac{1+z}{2})^{1/2}(\frac{\xi_{\rm e}^{2}}{0.1})(\frac{\xi_{\rm B}}{0.1})^{1/2}E_{52}^{1/2}t_{\rm day}^{-3/2}\,\,
{\rm eV}$, where $z$ is the source redshift. We see
that if we take $\xi_{\rm B}\sim 10^{-8}$ and $A_{\star}\sim 5$, then the critical frequency $\nu_{\rm c}$
may cross the optical band at about one day after the burst. On the other hand, we can
further calculate the peak frequency of ICS as
\be
\nu_{\rm n}=6.5\times 10^{5}(\frac{1+z}{2})(\frac{\xi_{\rm e}^{2}}{0.1})^{2}(\frac{\xi_{\rm B}}{0.1})^{1/2}A_{\star}^{-1/2}
E_{52}t_{\rm day}^{-2} \,\,\,\,{\rm eV}
\ee
Obviously, if we take the same parameters as above, then $\nu_{\rm n}$ should cross the optical 
band about 6 days after the burst.

The afterglow light curves will also be greatly modified by ICS. It has been shown that the 
electron Lorentz factor $\gamma _{\rm e}\propto t^{-1/4}$, the typical synchrotron radiation 
frequency $\nu_{\rm m}\propto t^{-3/2}$,  and the comoving specific intensity of synchrotron
radiation at peak frequency is $I_{\nu_{\rm m}}\propto t^{-5/4}$. From the relation $I_{\nu_{\rm n}}/
I_{\nu_{\rm m}}\sim R\gamma_{\rm e}^{-2}$, it is easy to show that the intensity of ICS at peak 
frequency is $I_{\nu_{\rm n}}\propto t^{-7/4}$, and $\nu_{\rm n}\propto t^{-2}$, then the observed peak
flux $F_{\nu_{\rm m}}\propto t^{-1/2}$, and $F_{\nu_{\rm n}}\propto t^{-1}$. Therefore, we conclude
that if our observation is fixed at frequency $\nu$, then the observed flux has four components:
$F_{\nu }\propto F_{\nu _{\rm m}}(\nu /\nu _{\rm m})^{-\alpha }\propto t^{-(1+3\alpha )/2}$ for $\nu <\nu _{\rm m}$,
$F_{\nu }\propto F_{\nu _{\rm m}}(\nu /\nu _{\rm m})^{-\beta }\propto t^{-(1+3\beta )/2}$ for $\nu_{\rm m}<
\nu <\nu _{\rm c}$, $F_{\nu }\propto F_{\nu _{\rm n}}(\nu /\nu _{\rm n})^{-\alpha }\propto t^{-(1+2\alpha )}$ for 
$\nu_{\rm c}<\nu <\nu _{\rm n}$, and $F_{\nu }\propto F_{\nu _{\rm n}}(\nu /\nu _{\rm n})^{-\beta }\propto 
t^{-(1+2\beta )}$ for 
$\nu >\nu _{\rm n}$. So we can see that, if take $\alpha =0$, $\beta =1$, then $F_{\nu }\propto t^{-1/2}$
for $\nu <\nu _{\rm m}$,  $F_{\nu }\propto t^{-2}$ for $\nu_{\rm m}<\nu <\nu _{\rm c}$,  $F_{\nu }\propto t^{-1}$
for $\nu_{\rm c}<\nu <\nu _{\rm n}$, and  $F_{\nu }\propto t^{-3}$ for $\nu >\nu _{\rm n}$.

\section{The effect of ICS in the non-relativistic case} 
For non-relativistic blast wave, $\hat {\gamma }=5/3$, and $\Gamma \sim 1$, then $R=
2\times10^{-3}(\frac{\xi _{\rm e}^{2}}{0.1})\beta ^{5}nt_{\rm day}$. We now consider two cases (the density 
$\rho \propto r^{-s}$) for both uniform density ($s=0$) and for wind density ($s=2$).

\subsection{s=0}  In this case, the evolution of blast wave velocity is $\beta =(t/t_{\rm NR})^{-3/5}$, where
$t_{\rm NR}\simeq 168(\frac{1+z}{2})E_{52}^{1/3}n^{-1/3}\,\,{\rm days}$. Then it is easy to show that $R(t_{\rm NR})
=0.34(\frac{\xi_{\rm e}^{2}}{0.1})E_{52}^{1/3}n^{2/3}$, and $R(t)=R(t_{\rm NR})(t/t_{\rm NR})^{-2}$.
So we see that in general case the ICS is unimportant unless the surrounding density is
very high. 

If the ICS is important,  then, as in the previous section,  we can write the typical quantities 
as follows: $\nu _{\rm m}(t_{\rm NR})=
2.8\times 10^{-5}(\frac{\xi _{\rm e}^{2}}{0.1})(\frac{\xi _{\rm B}}{0.1})^{1/2}n^{1/2}\,\,{\rm eV}$, and $\nu _{\rm m}(t)=
\nu _{\rm m}(t_{\rm NR})(t/t_{\rm NR})^{-3}$; for $\alpha =0$, $\beta =1$, $\nu _{\rm c}(t_{\rm NR})=
1.5(\frac{\xi _{\rm e}^{2}}{0.1})(\frac{\xi _{\rm B}}{0.1})^{1/2}E_{52}^{-1/3}n^{-1/6}\,\,{\rm eV}$, 
and $\nu _{\rm c}(t)=\nu _{\rm c}(t_{\rm NR})(t/t_{\rm NR})^
{-17/5}$; $\nu _{\rm n}(t_{\rm NR})=0.63(\frac{\xi _{\rm e}^{2}}{0.1})^{2}(\frac{\xi _{\rm B}}{0.1})^{1/2}n^{1/2}\,\,
{\rm eV}$, and
$\nu _{\rm n}(t)=\nu _{\rm n}(t_{\rm NR})(t/t_{\rm NR})^{-27/5}$. The specific intensity at peak frequency is
$I_{\nu _{\rm m}}\propto t^{-1/5}$, $I_{\nu _{\rm n}}\propto t^{1/5}$, and the observed peak flux
$F_{\nu _{\rm m}}\propto t^{3/5}$, $F_{\nu _{\rm n}}\propto t$. Therefore, the observed flux
at fixed frequency $\nu $ is $F\propto t^{3/5-3\alpha }$ for $\nu <\nu _{\rm m}$, $F\propto t^{3/5-3\beta }$ 
for $\nu _{\rm m}<\nu <\nu _{\rm c}$, $F\propto t^{-(27\alpha -5)/5}$ for $\nu _{\rm c}<\nu <\nu _{\rm n}$, and
$F\propto t^{-(27\beta -5)/5}$ for $\nu >\nu _{\rm n}$.

\subsection{s=2}  In this wind environment, Chevalier \& Li (1999b) and Wei \& Lu (1999b) have given 
the blast wave evolution $\beta =(t/t_{\rm NR})^{-1/3}$, where $t_{\rm NR}\simeq 1000E_{52}A_{\star}^{-1}$.
Then we can obtain $R(t_{\rm NR})=0.09(\frac{\xi _{\rm e}^{2}}{0.1})A_{\star}^{2}E_{52}^{-1}$, and 
$R(t)=R(t_{\rm NR})(t/t_{\rm NR})^{-2}$. It is also obvious that the ICS is usually unimportant unless
the mass loss rate is very high.

We can write the typical quantities as before: $\nu _{\rm m}(t_{\rm NR})=6\times 10^{-6}
(\frac{\xi _{\rm e}^{2}}{0.1})(\frac{\xi _{\rm B}}{0.1})^{1/2}A_{\star}^{3/2}E_{52}^{-1}\,\,{\rm eV}$, 
and $\nu _{\rm m}(t)=\nu _{\rm m}(t_{\rm NR})(t/t_{\rm NR})^{-7/3}$; 
for $\alpha =0$, $\beta =1$, $\nu _{\rm c}(t_{\rm NR})=1.25(\frac{\xi _{\rm e}^{2}}
{0.1})(\frac{\xi _{\rm B}}{0.1})^{1/2}A_{\star}^{-1/2}\,\,{\rm eV}$, and $\nu _{\rm c}(t)=\nu _{\rm c}
(t_{\rm NR})(t/t_{\rm NR})^{-5/3}$; $\nu _{\rm n}(t_{\rm NR})=0.14(\frac{\xi _{\rm e}^{2}}{0.1})^{2}
(\frac{\xi _{\rm B}}{0.1})^{1/2}A_{\star}^{3/2}E_{52}^{-1}\,\,{\rm eV}$, 
and $\nu _{\rm n}(t)=\nu _{n}(t_{\rm NR})(t/t_{\rm NR})^{-11/3}$. The specific intensity at peak frequency is
$I_{\nu _{\rm m}}\propto t^{-5/3}$, $I_{\nu _{\rm n}}\propto t^{-7/3}$, and the observed peak flux
$F_{\nu _{\rm m}}\propto t^{-1/3}$, $F_{\nu _{\rm n}}\propto t^{-1}$. Therefore, the observed flux
at fixed frequency $\nu $ is $F\propto t^{-(1+7\alpha )/3 }$ for $\nu <\nu _{\rm m}$, $F\propto t^{-(1+7\beta )/3 }$ 
for $\nu _{\rm m}<\nu <\nu _{\rm c}$, $F\propto t^{-(3+11\alpha )/3}$ for $\nu _{\rm c}<\nu <\nu _{\rm n}$, and
$F\propto t^{-(3+11\beta )/3}$ for $\nu >\nu _{\rm n}$.

\section{Discussion and conclusion}

The detection of GRB afterglows has greatly furthered our understanding these objects. In
particular, the shape of the afterglow light curves provides important imformation for exploring
their emission mechanism. Here we have calculated the effects of ICS on the GRB afterglows.
We have shown that, when the blast wave is relativistic, the ICS emission is usually important
for both uniform medium and wind environment, while when the blast wave is non-relativistic,
the effect of ICS is usually unimportant unless the surrounding medium density is very high,
such as $n\sim 10^{6}\,{\rm cm^{-3}}$ as proposed by Dai \& Lu (1999).

When the ICS contribution is important, then it may have great influence on the shape of
afterglow light curves, i.e.  it can flatten or steepen the light curves. In order to verify our analytical
results, we make a numerical calculation. Fig.1 gives our numerical results, where we take 
$\xi _{\rm e}\sim 0.3,\,\xi _{\rm B}\sim 10^{-7},\,A_{\star }\sim 10,\,\,p =3$. We see that although
the break is not as sharp as predicted, the flatten and steepen of the light curve is still obvious.

\begin{figure}
\centerline{\epsfig{file=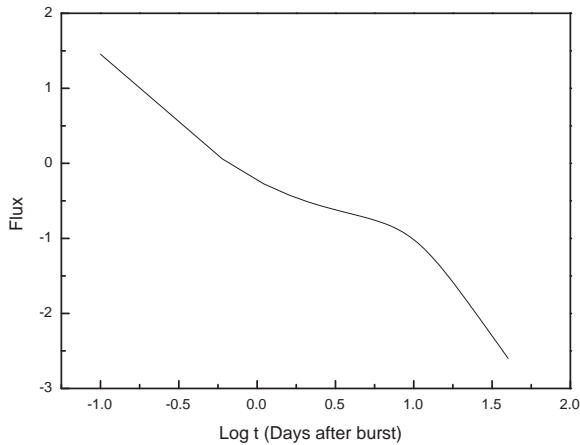, width=9cm}}
\caption{The possible shape of afterglow light curves when including the effects of ICS. }
\end{figure}

Although we have shown that the light curves of GRB afterglows may consist of four components
if ICS is important, it should also be noted that not all the GRB afterglows could have so many
components; the necessary condition for having four components is $\nu _{\rm c}<\nu _{\rm n}$, otherwise
there are only two components. The effect of ICS is strongly dependent on the model 
parameters, such as the energy fraction occupied by electrons, the total burst energy, the 
surrounding medium density, and the spectral index of synchrotron radiation. In particular,
the contribution of ICS is more important for those bursts with large spectral indices and 
dense medium.

Up to now there are total three bursts (GRB990123, GRB990510 and GRB000301c) for which  strong
breaks in their light curves are clearly observed. They are usually interpreted as evidence for 
collimation of the GRB ejecta (Rhoads 1999) or the transition from relativistic to non-relativistic
phase (Dai \& Lu 1999). Here we suggest that the effect of ICS can also produce the break 
in the afterglow light curves, which can either flatten or steepen the light curves, make the
temporal  behavior complicated.

\acknowledgements  We thank the referee for several important comments that improved this paper.
This work is supported by the National Natural Science Foundation
(19703003 and 19773007) and the National Climbing Project on Fundamental Researches
of China.


\begin{thebibliography}{}
\bibitem[]{} Castro-Tirado, A.J., et al., 1999, Science, 283, 2069
\bibitem[]{} Chevalier, R.A., Li, Z.Y., 1999a, ApJ, 520, L29
\bibitem[]{} Chevalier, R.A., Li, Z.Y., 1999b, astro-ph/9908272
\bibitem[]{} Dai, Z.G., Lu, T., 1998, MNRAS, 298, 87
\bibitem[]{} Dai, Z.G., Lu, T., 1999, ApJ, 519, L155
\bibitem[]{} Dermer, C.D., Boettcher, M., Chiang, J., 1999, astro-ph/9910472
\bibitem[]{} Fruchter, A., et al., 1999, ApJ, 516, 683
\bibitem[]{} Galama, T.J., et al., 1999, Nature, 398, 394
\bibitem[]{} Harrison, F.A., et al., 1999, ApJ, 523, 121
\bibitem[]{} Kulkarni, S.R., et al., 1999, Nature, 398, 389
\bibitem[]{} Kumar, P., Panaitescu, A., 2000, astro-ph/0003264
\bibitem[]{} Masetti, N., et al., 2000, astro-ph/0004186
\bibitem[]{} M\'{e}sz\'{a}ros, P., Rees, M.J., 1997, ApJ, 476, 232
\bibitem[]{} Moderski, R., Sikora, M., Bulik, T., 2000, ApJ, 529, 151
\bibitem[]{} Panaitescu, A., Kumar, P., 2000, astro-ph/0003246
\bibitem[]{} Panaitescu, A., M\'{e}sz\'{a}ros, P., 1999, ApJ, 526, 707
\bibitem[]{} Rhoads, J.E., 1999, ApJ, 525, 737
\bibitem[]{} Rhoads, J.E., Fruchter, A.S., 2000, astro-ph/0004057
\bibitem[]{} Sari, R., Esin, A., 2000, astro-ph/0005253
\bibitem[]{} Stanek, K.Z., et al., 1999, ApJ, 522, L39
\bibitem[]{} Veillet, C., Boer, M., 2000, GCNC  611
\bibitem[]{} Vietri, .M., 1997, ApJ, 478, L9
\bibitem[]{} Waxman, E., 1997a, ApJ, 485, L5
\bibitem[]{} Waxman, E., 1997b, ApJ, 489, L33
\bibitem[]{} Wei, D.M., Lu, T., 1998a, ApJ, 499, 754
\bibitem[]{} Wei, D.M., Lu, T., 1998b, ApJ, 505, 252
\bibitem[]{} Wei, D.M., Lu, T., 1999a, ApJ, accepted (astro-ph/9908273)
\bibitem[]{} Wei, D.M., Lu, T., 1999b, astro-ph/9912063
\bibitem[]{} Wijers, R.A.M.J., Rees, M.J., M\'{e}sz\'{a}ros, P., 1997, MNRAS, 288, L51
\end{thebibliography}
\end{document}